\documentclass[aps,twocolumn,twoside,floatfix,nofootinbib,a4paper]{revtex4}
\usepackage{graphicx}
\usepackage{marginnote}
\usepackage{bbm}
\usepackage{hyperref}
\usepackage{enumitem,kantlipsum}
\usepackage[usenames,dvipsnames]{color}
\usepackage{amsmath}
\usepackage{amssymb}
\usepackage{amsfonts}
\usepackage{mathrsfs}
\usepackage{bm}
\usepackage[all]{xy}
\usepackage{hyperref}
\usepackage{tikz-cd}

\newcommand{\beq}{\begin{equation}}
\newcommand{\eeq}{\end{equation}}
\newcommand{\bea}{\begin{eqnarray}}
\newcommand{\eea}{\end{eqnarray}}
	
\newcommand{\ea}{\end{align*}}

\newcommand{\bma}{\begin{pmatrix}}
\newcommand{\ema}{\end{pmatrix}}

\usepackage[percent]{overpic}
\usepackage{overpic}

\begin{document}
\title{A dynamical systems perspective on the celestial mechanical contribution to the emergence of life}
\author{Fan Zhang} 
\affiliation{Institute for Frontiers in Astronomy and Astrophysics, Beijing Normal University, Beijing 102206, China}
\affiliation{Gravitational Wave and Cosmology Laboratory, School of Physics and Astronomy, Beijing Normal University, Beijing 100875, China}
\affiliation{Advanced Institute of Natural Sciences, Beijing Normal University at Zhuhai, Zhuhai 519087, China}

\date{\today}

\begin{abstract}
\begin{center}
\begin{minipage}[c]{0.9\textwidth}
Biological activities are often seen entrained onto the day-night and other celestial mechanical cycles (e.g., seasonal and lunar), but studies on the origin of life have largely not accounted for such periodic external environmental variations. We argue that this may be an important omission, because the signature replication behaviour of life represents temporal memory in the dynamics of ecosystems, that signifies the absence of mixing properties (i.e., the dynamics are not fully chaotic), and entrainment onto regular, periodic external perturbative influences has been proven capable of suppressing chaos, and thus may bring otherwise unstable chemical reaction sets into viability, as precursors to abiogenesis. As well, external perturbations may be necessary to prevent an open dissipative (bio)chemical system from collapsing into the opposite extreme -- the point attractor of thermal equilibrium. In short, life may precariously rest on the edge of chaos, and open-loop periodic perturbation rooted in celestial mechanics (and should be simulated in laboratory experiments in origin-of-life studies) may help with the balancing. 
Such considerations, if pertinent, would also be consequential to exobiology, e.g., in regard to tidal-locking properties of potential host worlds.   
\end{minipage} 
 \end{center}
  \end{abstract}
\maketitle

\raggedbottom

\section{Ecosystem facet}
Evolved life-forms have not been observed to exhibit violent chaotic reconstruction across generations (we even see ancient species changing little over time), or else advantageous features may not have sufficient time to accumulate and manifest in population proportions, before being displaced by new alterations. Therefore, when viewed as a time series or dynamical process, successful evolution (including the initial chemical type\footnote{There has been increasing recognition (see e.g., \cite{Markert2015}) that the biological and chemical evolutions are just artificially separated terms referring to vaguely different segments of one continuous process.}) could plausibly occur in a context where there are regulating mechanisms at work, that reign in the otherwise nonlinear chaotic dynamics (cf., the population dynamics models, see e.g., \cite{Bacaer2011}), sufficiently to achieve quasi-stability\footnote{Compare with e.g., dynamic kinetic stability \cite{Pross2012WhatIL,Pross+2005+1905+1921}. We do not enlist kinetic barriers and competitive evolution into deeper potential wells for the attainment of meta-stability, but note multiple mechanisms may be at work simultaneously.}. 
More saliently, abiogenesis, from the perspective of dynamical systems, is the tendency for transient but repeating patterns (i.e., individual creatures or ``vortexes of life'' \cite{vortexlife}) to emerge and propagate. Or in other words, copies of quasi-stable (persistent for a lifespan) laminar phases emerge, and they keep multiplying, appearing in more phase space sub-dimensions, associated with additional copies of the chemical jigsaw pieces. This theme of persistence and repetition is antithesis to ergodic and mixing attributes\footnote{Statistical physics and thermodynamics describe systems possessing the mixing property (note, out of equilibrium thermodynamics is still thermodynamics, thus also requires this; e.g., increase of entropy is not guaranteed even in a probability sense, if the system trajectory in phase space does not transversely cover all volumes with equal density), but most systems, especially those exhibiting the colloquial ``complexity'' or ``emergent'' behaviour, such as biological entities, are not of this type. Biology in particular, exhibit so much heterogeneity in the form of different cells and organs all doing different things yet connected, that a statistical mechanical depiction must fall short, as one cannot hope to maintain any useful information after averaging/coarse graining.}, which entails a loss of memory of the initial conditions or earlier states, and can be regarded as useful statistical descriptions of chaos \cite{ergochaos,sep-ergodic-hierarchy,doi:10.1093/bjps/axn053} (the dynamical distribution of visiting frequencies over strange attractors provides the link with Ergodic Theory, see e.g., \cite{10.1063/1.1337757}; statistic depictions are useful since chaotic systems, even deterministic, would be practically unpredictable). Therefore, a necessary condition for replicating biology to arise, is the existence of suppression mechanisms that reign in the violence\footnote{That is, the number of positive Lyapunov exponents shouldn't be too numerous, and the amplitude of each shouldn't be too large.} of chaotic chemical evolvement\footnote{We adopt this phraseology to avoid muddling with the more restrictive Darwinian evolution.} in a multi-agent systems-chemistry soup, whose intrinsic guiding equations are nonlinear.    

On the other hand, not being too stable is also beneficial or indeed vital. If we have a very open system that is highly dissipative, leaking ``energy''\footnote{In the sense of a Liouville-theorem-style phase space volume. The Hamiltonian is one possibility of a specific such ``energy''.} quickly, then it will be stuck at the very stable lowest ``energy'' state, with any fluctuation and perturbation quickly damped out. Such a system would be rather inanimate, without metabolism, allowing only death for individuals and extinction for species. Indeed, premature thermal ``death'' seems to be the hurdle encountered in many real origin-of-life experiments (see e.g., \cite{Colomer2018ATS}). 

Both sides considered, seen from the perspective of chemistry, a life-supporting ecosystem contains a set of reactions that somehow is kept ongoing and away from a quiescent terminal state, but not via taking extreme swings and detours, which although might prolong the journey time, would also increase the chance of depleting some necessary resources, resulting also in extinctions. 
In other words, a mechanism needs to be present during abiogenesis that leaves ecosystems on the edge, possibly in a crevice between alternative definitions of chaos. Specifically, the evolvement would ultimately be aperiodic \cite{10.1063/1.1337757} as new species should emerge, yet mixing is not attained which would force amnesia of ancestors and the tricks of survival they learned. 

\section{Dynamical system facet}
In short, ``life'' seems inclined to rest on a fine edge between order and chaos. Such a precarious balance may be difficult to achieve, and we must look closely at what determines the severity of chaos. The first and most important criteria is nonlinearity, and we immediately notice that the rates of reactions are often nonlinearly related to the concentrations, because, e.g., when two reactants interact, the rate that depends on chance encounters of the two types of molecules is proportional to the product of both concentrations. 

The next obvious factor to look at is dimensionality. In particular, we recall that for very low (two\footnote{This refers only to the ``active'' ingredients whose concentration levels fluctuate during the evolvement; food or waste chemicals whose concentrations either don't change noticeably or don't matter to reaction rates are not counted. The very earliest steps in chemical evolution may well already involve more active species of chemicals, but as we discuss in this section, two, plus periodic external perturbations, represents the minimal requirement for nontrivial dynamics to ensure. As more diverse and complex initial setups may be quite depressed in turns of their probabilities of arising naturally by chance, we think this minimalistic scenario deserves careful scrutiny.}) dimensional deterministic systems described by differential equations, the requirement of the continuous trajectories to not cross each other, leading to ambiguity in the next step to take at the intersection point, means the trajectories cannot really zoom around in a messy chaotic manner, as the entire space is only one dimension larger than the trajectories themselves, so headroom is very limited. As a result, stable limit cycle or fixed point behaviors are observed instead \cite{wiggins1990introduction}. This is too far off the edge of chaos and into the stable territory, thus is not suitable for life\footnote{For completeness, we should mention a subtlety when it comes to chemical reactions. Consider a two-species interaction, because our initial detailed phase space is of an extremely high dimensionality (proportional to the number of molecules), the chemical reactions would almost certainly follow a chaotic trajectory there. However, if we project down to e.g., just the concentrations of the two chemical species as is common practice, then where does the chaos go? It would be quite unlikely that the projected trajectory just happens to become perfect limit cycles, so deterministic differential equation description of reaction dynamics, predicting continuous deterministic trajectories in the reduced concentrations space, must be missing something qualitative as well, beyond just quantitative accuracy. Indeed, the more realistic stochastic and discrete simulations of \cite{GILLESPIE1976403,doi:10.1021/j100540a008} (see in particular the Lotka and Brusselator reaction simulations in \cite{doi:10.1021/j100540a008}) do away with the restriction of non-intersection (both due to discrete jumping in location allowing trajectories to hop over each other, and stochasticity that removes the need for unique future evolvement -- quite suitably, indeterminacy exacerbates unpredictability), and therefore permit chaos in principle. Nevertheless, as the deterministic differential equation depiction of chemical reactions seems to be adequate in most applications (e.g., when jumpy effects due to small number statistics are not important), this subtlety may need rather long temporal durations to manifest itself in the origin-of-life chemical evolution context (cf., discretization also happens in finite-difference numerical solutions to the differential equations, and errors may build up only gradually). If this buildup period lasts longer than suitable conditions (e.g., liquid water) on a planet persist, then the subtlety may not by itself be sufficient to bring about dynamism (but there may also be that more than one factors are at work).}. Higher dimensionality would thus be preferred. 

To see whether high dimensionality is available, we begin by noting that, \emph{a priori}, the dimensionality of the phase space in our context should refer to the number of locations, velocities and other relevant configurational data\footnote{Like orientation, shape distortion etc; each described with a generalized coordinate in Hamiltonian mechanics.}, for each and all of the rudimentary chemical components related to biology (e.g., each of the food monomers, if the formation of polymers is being investigated). 
In practical computations however, almost all of the microscopic molecule-specific data are unknown, so their impacts are often subsumed into e.g., a stochastic averaging step \cite{GILLESPIE1976403,doi:10.1021/j100540a008}, leaving only macroscopic concentrations in the equations, essentially projecting out most of the dimensions in the phase space\footnote{\label{fn:laterlife0}Note we ignore spatial inhomogeneity, as we envision reagents all mixed in a puddle at the beginning of life, without cell walls. Cell walls, once developed, may also help curtail dissipation, so more sophisticated life-forms may rely less on periodic perturbation entrainments as chaos-inducement (cf., Footnote \ref{fn:laterlife} for chaos-suppression) mechanisms.}. 
The obvious way to get out of this quagmire of decimated dimensionality is to have more species of reagents interacting, but then the demand of having many bio-active molecules already present at the very beginning stages of chemical evolution may be quite onerous, and the resulting biochemical systems may in any case become too chaotic or even hyperchaotic\footnote{With more than one Lyapunov exponent being positive; larger number of dimensions allows a larger number of exponents, thus the chances for hyperchaos increases accordingly.} (see e.g., \cite{Filisetti2011ASM}), making a retreat onto the edge of chaos too difficult to attain with simple suppression mechanisms\footnote{\label{fn:laterlife}For later eons when life becomes more complex and spatially inhomogeneous, enzymes and energy/protein/genetic material production processes (especially the template-based replication) become so specialized that coupling between chemicals, i.e., admissible types of reactions, are severely severed, which in turn reduces the effective (Lyapunov \cite{1983JDE....49..185F} or information) dimension to render chaos more manageable. From this perspective, one might think Darwinian evolution as ``survival of the more stable'', and more advanced life-forms could thus have evolved to possess more sophisticated techniques for chaos-suppression (cf., Footnote \ref{fn:laterlife0} regarding chaos-inducement), escaping the need for the simpler periodic celestial mechanical entrainments, which we envision to be more important for the initial stages of abiogenesis. It is curious to note that theories of chaos control in low and high dimensional systems often employ different techniques (see e.g., \cite{scholl2008handbook,1995AdPhy..44...73S}), and Nature seems to have opted for the same.}. 

There is an alternative way out, namely, periodic external perturbations can come to the rescue. For example, periodic environmental parameter variations could imprint onto reaction rates, thus produce periodic parametric perturbations to the reaction dynamics. The now explicit time dependence present in the evolvement equations means that time has to be elevated into a new variable, so an otherwise two dimensional system becomes three dimensional, and chaos can ensue. On the other hand, the chaos thus introduced is hamstrung by the periodicity, as the evolvement within each period stays the same and so can be folded inside a Poincar\'e map \cite{Hirsch1974DifferentialED,Verhulst1989NonlinearDE}. In this way, the phase space can be factorized back down into the two dimensional Poincar\'e section again. However, this time, the evolvement via Poincar\'e map is discrete, so avoidance of intersections is no longer, and chaos is allowed, albeit in a more constricted capacity than full 3-D (e.g., the new Lyapunov exponent added by the periodic perturbation is zero \cite{1999PhRvE..59.5313M}, thus its direct impact is neutrally stable), possibly closer to the edge as we desire.  

Intuitively, one can understand this tendency to instill only restrained dynamism by viewing the external perturbation as injecting a moderate amount of ``energy''\footnote{Note, there is a subtlety, if the ``energy'' injection is so efficient that the system becomes conservative or even volume-growing, then even strange attractors cannot exist in the asymptotic temporal future, and nearby system trajectories could diverge wildly in all directions (in strange attractors, the trajectories diverge in some directions, but converge on others). However, we do not know what is the ultimate destiny of life, so constraints cannot be placed on the strictly mathematical asymptote. Instead, we can only assert that during the long but finite timespan when we observe the phenomenon of life, its biochemical evolvement should not be perturbed into being so wild, beyond even chaotic strange attractors. Therefore, we install a vague ``moderate'' qualification.}, but only preferentially into those spectrum components that are of resonant\footnote{We note ``resonance'' in the literature on chaos suppression by open-loop perturbation (e.g., \cite{1990PhRvA..41..726L,1994PhRvE..49..319K}) more restrictively refers to the commensurability between the frequencies of the perturbation and well-defined system  features such as the external driving forces that initiated the chaos.} (rational multiples \cite{1996CSF.....7.1555D,1999PhRvE..59.5313M}) frequencies. 
To see how this frequency-selectiveness reign in the wildness in the dynamics, we can begin with an already chaotic system\footnote{For easier exposition; similar behaviour is indeed, and perhaps even more cleanly, seen in bifurcating regimes leading up to chaos \cite{1986PhRvA..33..629W}.} and recall that unstable periodic orbits densely fill out its strange attractor \cite{1990PhRvL..64.1196O,scholl2008handbook}, so the spectrum frequencies can be defined here as the inverse of the periods of these limit cycles. Enhancing the power/weighting of a periodic orbit by injecting more ``energy'' into the corresponding frequency thus manifests as the system's phase space trajectory spending more time tracing near said orbit (see e.g., \cite{1999PhRvE..59.5313M}), in a protracted laminar flow phase. In other words, this orbit becomes more (but not necessarily fully) stable\footnote{A pictorial depiction of a structurally stable \cite{Arnold1983GeometricalMI} strange attractor would now appear more stratified, with lines associated with this orbit thickened. Structural instability, or larger perturbation amplitudes, may result in more drastic rewiring (see e.g., Footnote \ref{fn:Logi} below). Regardless, the dynamical distribution of visiting frequency to small neighbourhoods of the concerned cyclic orbit increases.} (as is shown quite explicitly by \cite{1991PhRvL..66.2545B}). Eventually the ``energy'' cascades into other frequencies via nonlinear couplings, but the leakage may not be rapid enough, as compared to the rates of injection and/or dissipation, to achieve a white-noise-like egalitarian spectrum, that signifies a Markovian-style path-independency (the evolvement during each time increment being independent to that during other increments)\footnote{In biological circumstances, if we suitably zoom in onto the ecosystem surrounding initially a single life-form, the laminar phase is when it is alive and metabolizing. After a while of the laminar flow, reproduction happens, and the orbit bifurcates into multiple copies of similar local flow patterns that exchange material (so they are constituents of a larger overall orbit). Eventually, the original local flow pattern disintegrates due to e.g., the ``energy'' leakage mentioned in the main text, but its copies persist elsewhere, and as long as reproduction outpaces disintegration, the overall orbit remains structured and manifestly path-dependent.}. In this case, the system does not come to die off at a fixed point, but is also capable of retaining past memory\footnote{One could also intuit how periodic entrainment would curtail chaos from another perspective, that of Bifurcation Theory. Recall that when we approach the chaotic regime by tuning system parameters, we typically witness a trend for the limit cycles to become more complicated (e.g., via period-doubling), until it finally loses periodicity altogether, in which case one could say that chaos has arrived \cite{10.1063/1.1337757}. The presence of one or more external reference periodicity, bleeding through into the dynamical evolvement via e.g., parametric perturbation, could re-enforce their resonant frequency windows in the bifurcation tree, and delay the onset of full chaos.}. 

It is worth noting that the use of open-loop (as opposed to involving feedback that needs adaptation to the system state) periodic perturbations to adjust chaos is a well established research arena, and we are merely advertising its importance in the origin-of-life context. We quickly recite a few concrete examples below to help hang our intuition on, and point to the references for interested readers. One group of examples illustrate how a periodic perturbation can introduce chaos, while the other on how it can also suppress its severity, therefore demonstrating that it has the capacity to attack on both fronts, providing a ``restoring force'' that keeps a dynamical system on the edge of chaos. The requirements on either side of the edge seem contradictory, but it has been noted explicitly in literature that, with slight adjustments, the same open-loop external periodic perturbation scheme can indeed achieve both chaos enhancement and suppression (see e.g., \cite{1996PhRvE..53..200C,1998IJBC....8.1693F}). This is not surprising, as complex systems tend to react to perturbing influences in delicate ways. It is therefore plausible that at some configurations of such perturbations, a fine balance between order and chaos, as we desire, is achievable with a single unified mechanism. 

Turning to the examples now, we recall that the study of the periodically perturbed damped pendulum \cite{10.1063/1.1337757} (see also e.g., \cite{1996PhRvE..53..200C,2001PhRvL..86.1737C}) belongs to the first group. Specifically, a pendulum becomes nonlinear when the swing amplitude becomes large, but a conservative system won't admit strange attractors. Adding dissipation on the other hand collapses the asymptotic state into a single attracting fixed point. To avoid this boredom, one could add a parametric perturbation, through varying the pendulum length, to inject ``energy'' and re-instate chaos. The chaotic motion thus obtained tends to be rather tame however, exhibiting a universal structure also seen in the simple \cite{NEWHOUSE2006672} H\'enon map. 

From the other side, chaos suppression is an intensely researched topic, with engineering applications in terms of removing the harmful consequences of unwieldy dynamics. Illuminating examples of open-loop perturbations-based strategies, our present context being an example of, are plentiful, some can be found in e.g., \cite{1987SPhD...32..270A,1991PhRvL..66.2545B,1993PhRvE..47.4585C,1993RuMaS..48..173L,1994Chaos...4..391L,1995PhRvE..51..761C,1998IJBC....8.1693F,2001PhRvL..86.1737C,2006Chaos..16b3109S,2005PhRvE..72c6206C,1999PhRvE..59.5313M,1996CSF.....7.1555D,1998PhRvE..58..423T,1996PhRvE..53..200C,1994PhRvE..49..319K,1990LNP...355..242P}, see also \cite{1995AdPhy..44...73S} and the many additional references within. A particularly familiar example of an external driver entraining an otherwise chaotic system onto its own periodicity is brought forth by simply rhythmically tapping on a leaky faucet \cite{WaterJet}. The droplets affect the pinching-off of later ones in complex ways, leading to chaotic dripping, but tapping will force premature dripping, at given fixed periodicity. Another familiar textbook example is when the pivot of an inverted/Kapitza pendulum is set in rapid vertical oscillation, in which case the pendulum becomes stabilized \cite{Arno89,1976168}.   

\section{Celestial mechanics facet}
In summary, an edge-of-chaos state could be attainable when a minimal\footnote{\label{fn:Logi}Chaotic systems are sensitive not only to initial conditions, but also to system/circumstantial/contextual parameters, and are thus pliable to control via external perturbations. For example, with Logistic map, the periodic windows densely fill the biotic potential range, so an infinitesimal alteration of that parameter will toggle the system between being chaotic and asymptotically periodic. Therefore subtle external perturbations, such as the celestial mechanical ones considered in this note, easily overlooked in origin-of-life studies thus far, can well exert seemingly disproportionate influences and alter the behaviour of (bio)chemical systems, that are characteristically complex.} amount of external perturbation is introduced. 
Indeed, the fact that our modern biological clocks, presumably corporealized by periodicities in our biochemistry, can easily synchronize with environmental or even sociological schedules, serves as an indirect empirical evidence, that similar entrainment of more primitive (bio)chemistry onto external perturbations may well also be possible. 
On pre-bio Earth, the simplest and most obvious environmental conditional variations are often tied to solar system celestial mechanics, some of which are:
\begin{enumerate}
\item The day-night cycle is directly related to the entropy extraction mechanism of Ref.~\cite{lifebook}\footnote{The presence of long-range forces such as gravity, breaks thermodynamics (cf., gravothermal catastrophe \cite{1980MNRAS.191..483L}; alternatively, simply note that the nonvanishing of mutual gravitational potential between two halves of a divided system means energy is no longer a simple extensive quantity, thus the sum function assumption \cite{khinchin1949mathematical} is violated), so the second law does not apply and entropy can very well decrease when astronomical processes are involved. Alternatively, one could attempt to define a generalized entropy that includes gravitational field contributions (see e.g., \cite{2010JPhCS.229a2013T}), so that the increase of the gravitational entropy offsets the decrease of the traditional statistical mechanical type, as signified by e.g., the appearance of life, and a generalized second law is thus preserved (note though, the non-mixing issue may mean that such an attempt at resurrecting thermodynamics, when biology is involved, might be inefficacious).}, that is postulated to sustain life at a more abstract level. This process prevents the attainment of thermal equilibrium, thus could be intimately related to, and indeed be elaborated by, the aforementioned ``energy'' injection functionality that we ask of the external perturber. 
Most tangibly, factors such as lighting and temperature variations, modulated by the day-night cycle, could most certainly instigate parametric perturbations that enter the reaction rates. 

\item The ebb and flow of lunar tidal strength alters the height of tides, thereby regulates the appearances of longer-lasting intertidal pools, which could be vital for e.g., concentrating chemicals to encourage polymerization \cite{1982JMolE..18..203D}, as the Sun dries out the pools gradually. See also e.g., \cite{1980Sci...209.1560G,1989S&T....78..452V,1991ccdm.book.....G,1991Icar...92..204C,1996UniCl..33....1C}, for other instances when the Moon's potential r\^oles within the origin-of-life context have been noted. 

\item For some more complex life-forms with sufficient lifespans at least, the seasonal changes strongly regulate the growth and reproductive processes, ergo terms such as ``the mating season''. As such, they must have been imbuing their sway into the evolvement dynamics of entire ecosystems. At the very earliest times, their influences may also be felt via e.g., freezing and unfreezing waterbodies.  

\end{enumerate}
That it taking no effort to come up with this highly abridged account of some possible entrainment channels, suggests that it is possible for regular periodic influences, of celestial mechanical origins, to insert themselves and become an indispensable part of the story of life, and so they need to be included or simulated in experimental origin-of-life studies. 

\section{Conclusion}
In this very brief note, we have argued that, much like how individual life-forms might perhaps operate on the edge of chaos (see e.g., \cite{TEUSCHER2022104693,2022arXiv221210595Z}), entire ecosystems, including the earliest ones hosting chemical evolutions, would likely also act the same. Such a delicate balance between chaos and order would demand mechanisms for its maintenance. For unsophisticated primordial pre-bio chemical goos, such mechanisms would likely have to be supplied extrinsically, and we have proposed periodic open-loop perturbation, according to celestial mechanical cycles within the solar system, as one possible candidate. We have unified the r\^oles of chaos-induction and suppression within the same mechanism, yet if the pre-bio reactions involve many reagents to begin with, the induction functionality would become unnecessary, and only the suppression aspect of our discussion remains applicable.  

The studies of complex systems are, suitably, quite complex, due in no small part to the variability between different dynamical systems, making reductive theoretical toy models rather ineffective as instruments of investigation. In particular, chaos-inducement and suppression strategies often lack rigorous mathematical proofs that they work in all situations. Yet so long as they possess intuitive/geometric appeal (e.g., \cite{1990PhRvL..64.1196O,1990LNP...355..242P}), they are often found to work in many experiments (see e.g., \cite{1995AdPhy..44...73S}). It is for this reason, and the fact that the details of the dynamical system, namely the reactions involved in abiogenesis, being lacking, that we have attempted to discuss the open-loop periodic perturbation mechanism in as many heuristic angles as we can master, much of it our own. 
We warn that, in order to provide more generally applicable intuition, that has a better chance of remaining relevant to the as-yet-unknown abiogenesis dynamics, we have left the discussion rather qualitative, broad-stroked and certainly not rigorous\footnote{At variance, more precise discussions regarding chaos control often rely on descriptive narrations of phase space structural features, that may be specific to the class of systems under investigation.}. 

We also caution that, the actual response of a system to attempted entrainment tends to be rather complicated, and desired regimes in the configuration space of possible perturbations may not be large \cite{1995AdPhy..44...73S,2006Chaos..16b3109S}. Fortunately, with biological systems, the reproductive feature means a small chance of success can quickly multiply into dominance\footnote{As a speculative example, the chiral preference shared by biomolecules could possibly be the result of most modern life-forms descending from a single ancestor, that just happened to choose between left and right handedness arbitrarily, without any deeper reasoning.}, and so the demand on robustness might not be too taxing. In other words, the problem that perturbation settings needing to be found via trial-and-error (especially for autonomous systems, for which an obvious preferred frequency, like that of a driving force, is absent), which hinders the application of open-loop chaos control strategies in engineering, is resolved by Nature through simply brute-forcing the many trials, possibly in the many ponds that host primordial chemical goos. 

Lastly, the lack of system-specifics also prevents us from commenting at all on the interplay between multiple external perturbers, such as that between the day-night, lunar and seasonal cycles, but this effect could well be pertinent in terms of providing the necessary flexibility to maintain chaos at the appropriate intensity for life (see e.g., \cite{2001PhRvL..86.1737C} for an example strategy of fine-tuning chaos with two perturbing terms, albeit at the same frequency). 

In any case, heuristics are no substitute for mathematical rigor and/or experimental verification, and more work will surely be required to assess the validity of our supposition. We speculate that origin-of-life experiments, for which simulated periodic perturbations can be added rather straight-forwardly (e.g., with ultraviolet lighting and temperature setting in periodic variation), can yield supporting or refuting evidences in the not-so-distant future. Alternatively, exoplanet observations, if ever passing some threshold so the detection of signs of (primitive) life-forms become routine, could provide the statistics that would allow us to assess the correlation with the availability of celestial mechanical periodic perturbers (e.g., planets tidally locked to their stars won't experience day-night cycles). 

\acknowledgements
This work is supported by the National Natural Science Foundation of China grants 12073005 and 12021003, and National Key Research and Development Program of China grant 2023YFC2205801. 

\bibliography{CelestialDriver.bbl}

\end{document}